\documentclass[12pt]{article}

\usepackage[a4paper,left=28mm,right=25mm,top=30mm,bottom=60mm]{geometry}
\usepackage[nonamebreak]{natbib} 

\newcommand{\ci}{\mbox{\protect{ $ \perp \hspace{-2.3ex}
\perp$ }}}
\newcommand{\n}[0]{\hspace*{.35em}}

\newcommand{\nn}[0]{\hspace*{.7em}}

\newcommand{\fourl}{\nn \nn}

\newcommand{\node}{\mbox {\LARGE
{$\mbox{$\circ$}$}}}

\newcommand{\snode}{\mbox {\large
{$\mbox{$\circ$}$}}}

\newcommand{\fla}{\mbox{$\hspace{.05em} \prec
\!\!\!\!\!\frac{\nn \nn}{\nn}$}}

\newcommand{\fra}{\mbox{$\hspace{.05em} \frac{\nn
\nn}{\nn
}\!\!\!\!\! \succ \! \hspace{.25ex}$}}

\newcommand{\dal}{\mbox{$  \frac{\n}{\n}
\frac{\; \,}{\;}  \frac{\n}{\n}$}}

\newcommand{\condnc}{\mbox{\raisebox{-.7ex}{\condnnc}}}
\newcommand{\condnnc}{\mbox {\LARGE{$\,
\mbox{$\Box$}
\raisebox{.21ex}{\hspace{-1.43ex}\mbox{$\circ$}}
$}} }

\newcommand{\margn}{\mbox {\raisebox{-.1 ex}{\margnn}}}
\newcommand{\margnn}{\mbox {\Large
{$\not \: \not $}}$\node $}

\newcommand{\flar}{\mbox{\raisebox{.5ex}{\fla}}}
\newcommand{\dall}{\mbox{\raisebox{-.5ex}{\dal}}}
\newcommand{\fladal}{\mbox{\flar \hspace{-1.82em}\dall}}

\newcommand{\pcal}{\ensuremath{\mathcal{P}}}

\usepackage{latexsym}
\usepackage{graphicx}

\RequirePackage{amsmath,amssymb}
\usepackage{here}

\date{\n} \vspace{-4cm}

\begin{document}

\noindent{In: (Wright, ed.) International Encyclopedia of the Social and Behavioral Sciences, 2nd ed., \textbf{10},  Elesevier, Oxford, 341--350.
 
\baselineskip=20pt

\begin{center} {\bf \Large Graphical   Markov Models: Overview\\} \end{center}
\begin{center}  {\large Nanny Wermuth* and  D.R. Cox**} \\[5mm]   \end{center}

{\em *Mathematical Statistics, Chalmers University of Technology, Gothenburg, Sweden and
Medical Psychology \& Medical Sociology,  Gutenberg-University, Mainz, Germany\\
e-mail: wermuth@chalmers.se, and **Nuffield College, Oxford University, Oxford, UK\\
e-mail: david.cox@nuffield.ox.ac.uk}\\[-2mm]

\noindent{\bf Abstract}
We describe  how graphical Markov models  emerged   
in the last 40 years, based on three essential concepts that had 
been developed  independently more than a century ago.   
Sequences of joint or single regressions and their regression graphs
are singled out as being the subclass that is best suited for analyzing
longitudinal data and for  tracing developmental 
pathways, both in observational and in intervention studies. Interpretations are illustrated using  two sets of data. Furthermore, 
some of the more recent, important  results for sequences of regressions are 
summarized.\\[-4mm]

\subsubsection*{1 Some general and historical remarks on the types of model}

Graphical 
models aim to describe  in  concise
form  the possibly complex interrelations between a set of variables so that  key properties 
can be read directly off a graph. The central idea is that each variable is
represented by a node in a graph. Any pair of nodes may become coupled, that is joined by an edge.
Coupled nodes are also said to be adjacent.
For many types of graph, a missing edge represents
some form of conditional independence between the pair of variables and an edge present 
can  be interpreted as a corresponding conditional dependence. 
Because the conditioning set may be empty, or  may contain some or all of the other variables,
 a variety of types of graph have been developed and are used to represent different types of
 structure. 

A particularly important distinction is between directed and undirected edges. 
In the former an arrow indicates the direction of dependence of a response on  an explanatory variable, the latter is also 
called a regressor.  If,  on the other hand,  two variables are to be interpreted on an equal standing then the  edge
between them is typically undirected.  For instance, systolic and diastolic blood pressure are   
treated as being on equal standing because they are two aspects of a single phenomenon, 
namely  a blood pressure wave.

Graphical Markov models started to be developed after 1970 as  special 
subclasses of log-linear models for  contingency tables and  of joint Gaussian distributions, where conditional independence constraints are imposed such that conditioning 
is on all the  other variables;  see Darroch et al. (1980), Wermuth (1976, 1980).
These models are  typically represented 
by undirected graphs with  edges that are full lines, called nowadays concentration graphs. The  same types of  graph were used  by the physicist
Gibbs (1902) to describe for two systems of particles having the same node sets,  one as more complex  whenever its nodes have  
more edges, that is larger numbers of  `nearest
neighbors'. 

 The first extension was to situations in which the variables can be arranged recursively, that is in an  ordered
 sequence, so that each variable turns into a single response to variables in its past and may be explanatory only
 to other variables in its future; see Wermuth (1980), Wermuth \& Lauritzen (1983). This led for single responses to  what
are now called  directed acyclic graphs, for sequences of  joint or single responses to the so-called regression graphs, and  to distributions said to be generated over graphs. 

For such generated Gaussian distributions, these models include as a subclass  the path analyses  of the geneticist Wright (1923, 1934). 
Wright had studied,  for his data,  sequences of exclusively linear regressions as possible stepwise generating 
processes. In his graphs, each missing arrow corresponds to  the vanishing of the partial correlation coefficient 
given all remaining variables  in the past of a given response, which   in Gaussian distributions is
a conditional independence constraint.  For other types of generated distributions,  an important issue is to identify 
parametric consequences of conditional independences; as for instance for the  so-called CG-regressions, see
Lauritzen \& Wermuth (1989), Edwards \& Lauritzen (2001).

 In general, the vanishing of a correlation coefficient may  coexist with a  nonlinear dependence. 
 Conversely,  a substantial partial correlation can occur in spite of conditional independence; 
  for an example,  see Wermuth \&
 Cox (1998). 
 Currently, one knows also for  jointly symmetric binary variables, generated over a
 directed acyclic graph, that an arrow vanishes  in the graph if and only if there is  a corresponding zero partial 
 correlation; see Wermuth, Cox \& Marchetti (2009). Zero partial correlations given 
 all other variables may reflect conditional independences  more generally, provided the distribution 
 is  generated over  even more specialized types of  graph;  see Loh \& Wainwright (2013), Wermuth, Marchetti
 \& Zwiernik (2014), Wermuth \& Marchetti (2014).

  The next extensions were to variables of any type 
 so that a missing edge  corresponds to a conditional independence. This exploits 
 a proposal by the probabilist Markov (1912):  many types of  seemingly complex joint distribution may be 
 much  simplified by  conditional independences. Important issues are  
 defining sets of independences  for a graph and  criteria to derive all  
implied  independences  so that an  independence structure captured by the  graph becomes well 
understood; see here Section 4.

 Further developments include different types of chain graph models; see Cox \& Wermuth (1993), Drton (2009). 
 The name `chain graph'  reflects  a full ordering of the variables
 into a sequence of joint or single responses and possibly a  set of  remaining variables that is to
 capture the context of a study or  properties of individuals  given at the baseline and hence regarded as given. 
 For  parameterizations  of Gaussian distributions with different types of chain graph, 
 see Wermuth, Wiedenbeck \& Cox (2006).
  
 Statistical monographs documenting the  early development of what are now
called graphical Markov models are  by Whittaker (1990), Edwards (1995), Lauritzen
(1996), Cox \& Wermuth (1996). For  surveys of probabilistic aspects,  see 
Pearl (1988) and Studen\'{y}  (2005). For graphical models used  in 
 expert systems, see  Cowell et al. (1999) and  for machine learning, see 
Wainwright \& Jordan (2008). 

Special  theoretical results, not touched upon here, have been derived under the assumption that a given
probability distribution satisfies all independence constraints represented by a 
graph and no more. Such distributions have been said to be  `faithful to the 
graph'; see Spirtes, Glymour \& Scheines (1993).

The existence of a family of  faithful distributions means that one can  choose a member at random, constrained just by the independences specified by a graph, and this  distribution captures precisely the independence structure of the graph. 
But  such an existence result is  of no relevance when other constraints are known for 
subsets of parameters. For instance, the whole family of Gaussian distributions  is  faithful to concentration graphs,  but  there is a simple  subfamily with additional parameter constraints that is  always `unfaithful' to concentration graphs, see Wermuth (2012), Section 2.5.

More importantly,  for data analyses, in which it is crucial to respect known important dependencies, such an existence is also of  no 
relevance.  In most research settings, much is known about subsets of parameters because  the direction and strength  of some dependences derive from substantive theory  or from results in previous  empirical studies.

 \subsubsection*{2 Sequences of regressions and data generating processes}

Directed acyclic graphs are  a common subclass of all
currently known types of chain graph. Distributions generated over a directed acyclic graph
 result  in a stepwise fashion in terms  of univariate conditional distributions, also called   recursive 
 single-response regressions.  These may  be compatible  with causal interpretations and therefore
appear attractive, at first sight,  for  substantive research that is driven by causal hypotheses.

 It is however well understood, from randomized clinical trials in particular, that real 
interventions often result in modifying more than a single response and possibly other features. For instance,
a medication that is to reduce blood pressure, will typically affect both systolic and diastolic pressure. Hence 
models that are to be compatible with causal interpretations should 
contain response variables that may respond at the same time to a change in relevant 
explanatory  variables. This is possible with joint response regressions.

 Nevertheless,  much of the current literature 
on causal modeling is based on directed acyclic graphs and  on only virtual  
interventions; see Pearl (2009, 2014 in press). With such a virtual intervention, those
changes in  single response  are recorded that result by fixing the values of some 
variables.  This may for instance be  treatments. One removes arrows pointing to these variables in a given graph and estimates treatment effects given these additional independence constraints.  Implicit is the assumption that a given  joint 
family of distribution remains unchanged,  otherwise. 

These recorded changes in virtual interventions, even though they are often  called `causal effects',  may  tell next to nothing about actual effects in  real 
interventions with, for instance, completely randomized allocation of patients to
treatments.  In such studies, independences  result by design and  they lead to missing arrows in well-fitting   graphs; see for exampleFigure 9 below, in the last  subsection.

 Sequences of regressions   in joint and single responses are the  most attractive chain graph models for observational 
 as well as intervention studies. For these models, 
residual associations among the responses may often be regarded as secondary features. But,  
standard univariate estimation methods for regression coefficients may lead to distorted estimates 
whenever two associated joint  responses on equal standing have disjoint subsets of regressors, a situation
that has become known as seemingly unrelated regressions.  When strong residual dependences
remain after one has been regressing each  response component separately on  two different sets of  regressors,
then distorted parameter estimates may result,
see  Haavelmo (1943), Zellner (1963), Drton \& Richardson (2004).

 For  known types of
chain graph other than regression graphs, recursive generating processes may not exist since   response components are conditioned  
on other joint responses on equal standing. 
While substantive theory may occasionally support such models,  this type of conditioning is typically
counterintuitive whenever one main aim of an analysis is to identify consequences of interventions or early predictors for  joint responses.
   But now, graphical criteria are available to decide whether the regression graph of a given sequence of joint or single regressions,
     defines the same independence structure as another chain graph, that is whether 
   they are Markov equivalent; see Wermuth \& Sadeghi (2012), Section 7, and here the subsection: Markov 
    Equivalence of Regression Graphs. 
       
   For applications, arguably  the most important  developments concern conditions under which the regression graph 
   can be used to trace development, to predict implied dependences in addition to implied  independences, as well as 
   graphical criteria  for possible confounding effects   when some variables are 
   ignored or subpopulations are studied. 

\subsubsection*{3 Regression graphs as partial summaries of data analyses} 

We give here two examples of data for which corresponding detailed statistical analyses have been published;
see the Appendix in Wermuth \& Sadeghi (2012) for the first and the Appendix  in Wermuth \& Cox (2013) for the second 
set of data. 
 
In  the regression graphs derived for both examples, the dashed lines capture remaining associations among 
 two joint responses and full lines, for  the context variables, reflect that conditioning is on all 
the context variables simultaneously. Estimation is  for both data sets by local modeling of each response alone on its past and 
of response pairs on the union of their directly explanatory variables. When there are only independence constraints concerning responses and no seemingly unrelated regressions, local  maximum-likelihood (m.l) estimates lead directly to proper   m.l.\,estimates for the joint distribution; see Cox (2006) for such types of  general principle.

 In both following examples, data analyses lead to traceable regressions; see Wermuth (2012). This means that for a given  ordered sequence of single and joint responses, the resulting  regression graph permits one not only to read off  independences, but more importantly for most applications, sequences of  directed edges 
show pathways of development. 

 Tracing of paths  is always possible for  joint Gaussian distributions whenever  each edge in its regression graph represents a substantial conditional  dependence; see here Section 4. This may not be possible with graphs of some structural equation models; see Wermuth 
 (1992). More importantly,  sufficient conditions for traceability using just the graph, are now known for other types of  distributions generated over regression graphs.

\subsubsection*{Data on chronic pain treatment\vspace{-2mm}}

For some medical data,  Figure 1 shows a first ordering of the variables. This ordering  was derived in discussions by physicians and statisticians, prior to the analyses.

We
are grateful to Judith Kappesser, now  Department of Psychology, Liebig-University of Giessen, for letting us use her 
data. They are for 201 chronic pain patients who have been given a three-week stationary treatment at a chronic pain clinic.
Two main research question are:  which development is most influential for success or failure of treatment and  is it necessary 
to include information on the
patient's site of main pain?

 The response of primary
interest is self-reported success of treatment, measured
three months after discharge by a score summarizing several aspects of the illness.  Among the context variables, capturing features that cannot be modified, were  age, gender,  lower or higher level of formal schooling and the number of
previous other illnesses. Only those are shown in Figure 1 that had sizeable dependences to the responses of
main interest.

 \begin{figure}[H]
\begin{center}
\includegraphics[scale=.56]{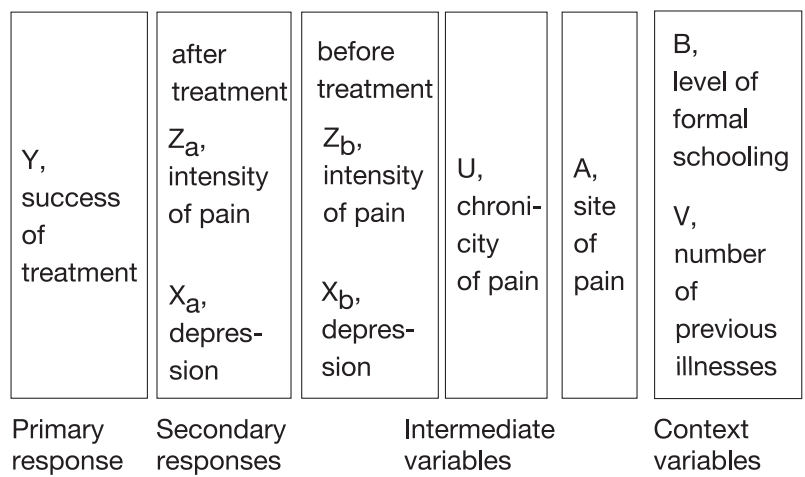}
\end{center}
\caption{\small Inititial ordering of variables in the chronic pain study for a sequence of regressions.
Variables within a same box are treated on an equal standing. Variables in any one box are considered conditionally 
given all variables their past, listed within boxes  to their right. } \label{fig1}\vspace{-5mm}
\end{figure}

There are a number of intermediate variables.  Before and  after a three-week stationary
treatment, questionnaire scores are available  of depression and  of intensity of   pain ranging from `no pain'
to `pain as strong as imaginable'. The chronification score incorporates different aspects, such as time since onset  of pain,
spreading of pain, use of pain relievers,  the patient's pain treatment history. Main site of
pain has here two categories: `back pain' and pain on the `head, face, or neck'. 

The regression graph of Figure 2 summarizes some aspects of the statistical  analyses.  Throughout, symbol $\ci$  means independence, symbol $\pitchfork$ means dependence and symbol $|$ is to be read as  `given'.
The graph shows, in particular, which of the
variables are needed so that for any given response, adding one more of the potentially explanatory variables  does not improve
prediction. Site of pain is an important intermediate variable since it is a node along a direction-preserving path of arrows
pointing from level of schooling to treatment success,
 hence should  be part of any future  study
of  chronic pain.

Some of the directions and type of dependences, which cannot be read off the graph, are as follows. Patients with many
years of formal schooling (13 years or more) are more likely to be head-pain patients, the others are more likely to be
back-pain patients, possibly because more of them have jobs involving heavy  physical work.  Back-pain patients  have
 higher scores  of pain  chronicity, reach  higher stages  of intensity of pain before treatment and report
higher intensity of pain after treatment.

\begin{figure}[H]
\begin{center}
\includegraphics[scale=.56]{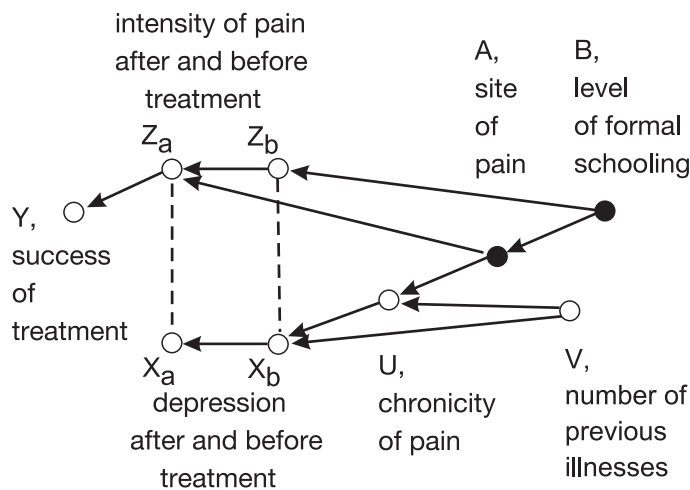}
\end{center} \vspace{-4mm}
\caption{\small Well-fitting sequences of regressions which  have a statistically  significant relation
for each  edge present in the graph.  Discrete  variables
are drawn as dots, continuous ones as circles. For instance, the two context variables are marginally independent, written as $B \ci V$ and
response  $Z_b$ is conditionally independent of $V$ given $A,B,U$;   $Z_b\ci V|A,\!B,\!U$, also $A$ is marginally dependent
on $B$; written as $A\pitchfork B$,  while $U$ depends on $V$ given $A$; $U\pitchfork V|A$.
{\bf Note}: Full arrows in regression graphs are sometimes also drawn as dashed-line arrows and  dashed lines as so-called arcs, which are full lines with arrow-heads at both ends} \vspace{-4mm}
\end{figure}

 Never captured by the graph alone are  nonlinear dependences, here  of  $Y$ on $Z_a$. Treatment success, $Y$, is  low whenever   higher levels of 
intensity of pain  remain after treatment, $Z_a$. But at  relatively lower levels of $Z_a$, treatment 
is  clearly the more successful the lower the intensity of pain at discharge, $Z_a$. 
The model  fits the data well since for each response taken separately, no indication was found that adding variables, further  nonlinear or interactive effects  would improve prediction and no strong dependences remained among the residuals of the joint responses on equal standing.

One important path of development is  that patients with shorter formal schooling are more likely 
to get chronic  back-pain and patients with chronic back-pain get help too late and 
respond less well to the type of treatment offered in the chronic pain clinic. This suggests  as possible interventions to modify the type of  treatment
for the back-pain patients or, much more ambitiously, to raise the general level of formal schooling.
 
\subsubsection*{Data on child development\vspace{-2mm}}

Here we use
for data  of 347 families participating in the `Mannheim study of children at risk'. 
We are grateful to Manfred Laucht, Central Institute of Mental Health, Mannheim, for  permitting a reanalysis of the data.

The study started with a random sample of more than 100 newborns from the general population of children born near Mannheim, Germany. This sample was completed to give roughly equal subsamples, in each of nine level combinations of the two types of adversity at birth, categorized  to be at levels `no, moderate or high'; see Laucht, Esser \& Schmidt (1997). In other words, there was heavy oversampling of children at risk for motoric or cognitive deficits in later years and   a random sample  served as a control group, which provides, in particular, comparable norm values.

The recruitment  of families  stopped with  362 children. 
All measurements were reported in
standardized form using  the mean and standard deviation of the starting random sample.  Of  the 362 German-speaking families who
entered the study when their first, single child was born without malformations or any other severe handicap,  347 families still participated when their child reached the age of 8 years. 

There are joint responses of main and of secondary interest at age 8 and at age 4 and a half  years. Each of these
contains two components of possible deficits: cognitive or motoric; see the derived graph in  Figure \ref{fig3}.

   \begin{figure} [H]
\begin{center}
\includegraphics[scale=.56]{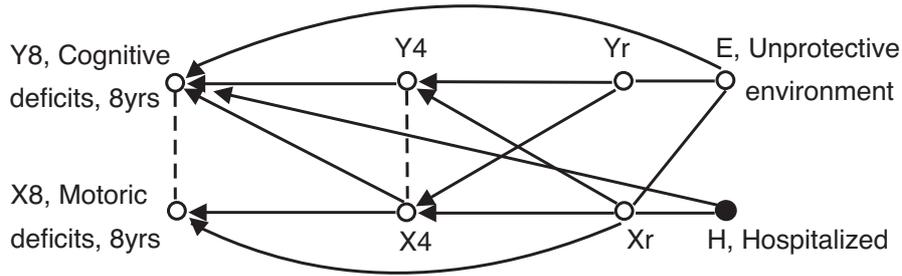}
\end{center}
 \caption{\small A well-fitting regression graph in the child development  study;
arrows point from regressors in the past to responses in the future; dashed lines are for dependent responses given their past; full lines are for dependent early risk factors given the remaining context variables.}\n \\[-4mm]
$\fourl \fourl \fourl \fourl  \fourl Y_8:\n  Y_4+X_4^2+E+H$  \fourl $X_8:\n X_4^2+X_r$ \\[1mm]
$\fourl \fourl \fourl \fourl  \fourl Y_4:\n Y_r+X_r^2$ \fourl  \fourl  \fourl \fourl $X_4: \n Y_r+ X_r^2$\\[1mm]
$\fourl \fourl \fourl \fourl  \fourl Y_r:\n E^2$\fourl  \fourl  \fourl  \fourl \fourl \fourl  $X_r:\n  E+H$\\[1mm]
 {\bf  \small Note:} \small{In this notation supplementing the graph, every square term implies that also 
a main effect is included in a regression;  see discussions of Wilkinson's 
    notation in McCullagh and Nelder (1983).}
 \label{fig3}  \end{figure}
 
  One main difference to previous analyses  is that we averaged three different
assessments of each of two types of risk: recorded at birth, at 3 months and at two years.  In both cases,   this can be justified by the six  observed  
pairwise correlations being all nearly equal; see e.g. Wermuth (2013). 
These averaged scores are risks  at two years for motoric deficits, labelled $X_r$, and  for cognitive  deficits, labelled $Y_r$. Two possible additional risks were identified at three months after birth,
 a score  called unprotective environment, $E$, and a binary variable $H$, which records whether the child had to be hospitalized during
 the first three months after birth.
 
 The graph shows in  particular  for motoric deficits at age eight years of the child,  $X_8$, that no arrows are pointing to it from  risks at 3 months, $E,H$, or  from 
 any variable for the cognitive side, $Y_8, Y_4, Y_r,$ given  information on the more recent motoric deficit, $X_4$ and the risk for motoric deficits, $X_r$.
 By contrast,  cognitive deficits at age 8 years, $Y_8$  depend directly on unprotective environment at 3 months,  on being hospitalized up to 3 months, as well as on motoric deficits at 2 years,  given both more recent deficits, $Y_4, X_4$.  A tentative  interpretation is 
 given  using Figure 
 \ref{fig6} in the next to 
    last subsection.
 
 The coding of variables implies  that all dependences are positive so that
 deficits accumulate with each additional regressor. Some of the effects are accelerated compared to linear dependences. This is not reflected in the
 graph alone, but in Wilkinson's notation which adds to the graph sums of highest-order
    effects; see legend of Figure \ref{fig3}.  
 
\subsubsection*{4 Questions regarding applications and statistical research}

Here, we first list questions  that  arise in general when new statistical models  are applied.  We  then give partial answers and references for sequences of  regressions.
\begin{description}
\item
$(i) $ Are studies available 
in  which the new  models have been used fruitfully?
\item
$(ii)$ Can well-fitting  models be derived from data and be tested as hypotheses formulated for  future data?  
\item
$(iii)$  What additional possibilities do the models offer to gain
better insight into given research questions, especially  to an understanding of development over time? 

\end{description}

As to $(i)$, case studies with sequences of regressions, which include joint responses, continue to accumulate. In addition to the two examples
summarized here in the previous section; see for instance Cox \& Wermuth (1996), Cheung \& Andersen (2003), Hardt et al. (2004),
Smith (2009), Hardt, Herke \& Schier (2011), Marchetti et al. (2011),  Schier et al. (2014 in press), Solis-Trapala et al. (2014 submitted).

As to $(ii)$, one most attractive feature of sequences of regressions is that their fitting requires typically no new estimation techniques. Standard methods
are often available;
for linear regressions, see Weisberg (2014),  for catego\-rical responses, see Cox (1972), 
 McCullagh \& Nelder (1989),  Andersen \& Skovgaard (2010).  To screen for nonlinear or interactive effects, see Cox \& Wermuth (1994). 
Regression graphs represent hypotheses about generating processes which can be tested in future studies.

When seemingly unrelated regressions are hypothesized, direct estimation methods are based for discrete variables  on generalized 
linear models; see Marchetti and Lupparelli (2011), for  joint  Gaussian distributions, estimation is possible with  structural equation models; see e.g. Bollen (1989).  But as mentioned before, when strong residual dependences remain 
in Gaussian seemingly unrelated regressions, especially for small sample sizes,  estimates may be far from  the population values.

One may also regard seemingly unrelated regressions  as so-called reduced models and embed them in  larger so-called covering
 models for which  estimation is again standard; see Cox \& Wermuth (1990).  Often m.l.\,estimates in the covering model are not far
 from those obtained by separate regressions and are then good approximations 
 to the m.l.\,estimates  in the reduced model.

As to  $(iii)$,  a number of new results have been obtained in the last years. We concentrate on the following few, discussed in the next three sections:
\begin{description}
\item
a) graphical conditions for the Markov equivalence of two regression graphs, to be used  for possible alternative interpretations or for
different types of fitting algorithm,
\item
b) conditions and criteria for deriving implied independences  and implied dependences from a given regression graph,

\item
c) ways of  distinguishing  different types of confounding that  are implied by a generating process when some variables are unobserved, 
that is marginalized over, or a population subgroup
is studied, that is when  there is conditioning on  the levels of some variables present in the generating process.
\end{description}

\subsubsection*{Markov equivalence of regression graphs\vspace{-2mm}}
Two different graphs are  Markov equivalent if they define the same independence structure. The independence structure of a regression graph, with  a given set of nodes and an ordering for the  responses, is determined by  its list of  missing edges for node pairs $i,k$.  Such a graph has possibly two types of undirected  edge, full lines between context nodes 
 and dashed lines between responses on equal standing, as well as arrows starting at a regressor and pointing to a response; see Figure \ref{fig1} to recall the meaning of these terms.

A missing full $i,k$-line between two context nodes means conditional independence given all remaining context nodes  and a 
missing dashed $i,k$-line between two response nodes on equal standing means conditional independence given other nodes in the past of $i$ and $k$. If a possible $i,k$-arrow pointing from $k$ to $i$ is missing, this means conditional independence given all other nodes in the past of $i$ except for $k$.

A  {\sf V}-configuration in such a graph,  often called just a {\sf V},  
consists of  three nodes with  two uncoupled nodes, called the outer nodes, $i,k$. Both outer nodes are coupled to 
 an inner node, \snode.  There are two types of {\sf V}s in regression graphs, named collision and transmitting {\sf V}s.
In each of the three  possible  types of collision {\sf V}s: 
$$ i \dal \snode \fla k,  \nn \nn  i  \fra \snode \fla k, \nn \nn i\dal \snode \dal k \, ,$$
 the two  edges are removed when the inner node
is marginalized over. In each of the remaining  transmitting  {\sf V}s, 
both edges are removed
    by conditioning on the inner node.

Directed acyclic graphs consist exclusively  of arrows so that there can be only the middle {\sf V} above, where two uncoupled regressors point to a common response.
Graphs of only full lines, often called concentration graphs, have only transmitting {\sf V}s, while graphs of only dashed lines, often called covariance graphs,  have only   collision {\sf V}s.

These three types of graph are subclasses of regression graphs. The names of the last two derive from parameters  in joint Gaussian distributions. For such a distribution, zero off-diagonal elements, $\sigma_{ik}$, in the covariance matrix capture a marginal independence, and zero off-diagonal elements, $\sigma^{ik}$, in the concentration  matrix, the inverse of  the covariance matrix, 
capture a conditional independence given all the remaining nodes.

Two different types of regression graph, which have the same node set and the same set of missing edges but different types of edge for some node pairs, are Markov equivalent  if and only if
their sets of collision {\sf V}s coincide, even though for any  given {\sf V}, there may be any of the three types of collision {\sf V}s above; for a proof, see Wermuth and Sadeghi (2012). One example is shown in Figure  \ref{fig4}.
\begin{figure} [H]
\centering   \vspace{-2mm} \includegraphics[scale=.56]{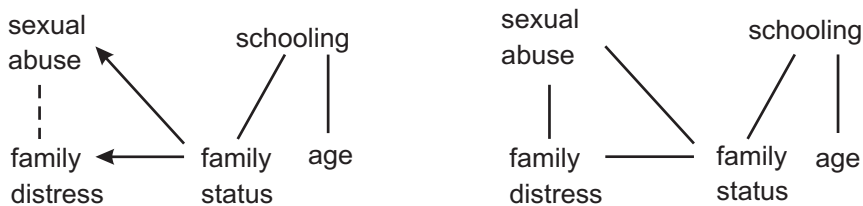}
 \caption[]{\small{A  regression graph that is Markov equivalent to a concentration graph.}}  \label{fig4}\vspace{-2mm}
 \end{figure}
 
 The graph on the left of Figure  \ref{fig4},  resulted from a retrospective study, with more variables than reported here. Questions about their childhood were answered by 283  adult  females when visiting a general practitioner  for some minor health problems; see Hardt (2008). No nonlinear or interactive effects were detected.
 
The  well-fitting graph  contains two binary variables, level of formal schooling and severe sexual abuse during childhood, as well as three quantitative measurements.  Age in years is recorded directly, the other variables being derived  from questionnaires.  Family status indicates  the recalled social standing of the family during  early childhood. Family distress includes psychological disturbances and  alcohol or
drug problems of the parents. 
 
 From  the Markov equivalence to  the concentration graph on the right of  Figure  \ref{fig4},
  one knows, for instance,  directly that  sexual abuse  is independent of the 
  level of formal schooling given knowledge about the family status. This implication may be derived  from the defining independences of the graph on the left or be based on the following separation result in graph theory. 
   If in an undirected graph with 
 three  disjoint subsets $\alpha,\beta,c$ of the given set of nodes, every path between $\alpha$  and $\beta$ has a node in $c$, then $\alpha$ is  separated  from $\beta$ by $c$, that is removal of set $c$ leaves $\alpha$ and $\beta$ disconnected.
  For concentration graphs, this separation  implies  $\alpha \ci \beta|c$; see e.g. Lauritzen  et al. (1990).
  
   Figure \ref{fig5} shows instead a regression graph on the left that is Markov equivalent to a covariance graph on the right. It is the simplest case 
   of the seemingly unrelated regressions with regressors labelled 3, 4 and  responses labelled  1, 2. For instance, it follows from the graph definitions that  edge $1\dal2$
   means $1\pitchfork 2|3,4$ on the left and $ 1\pitchfork 2$ on the right.

    \begin{figure} [H]
\centering    \includegraphics[scale=.5]{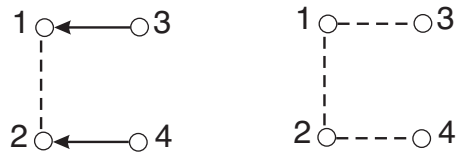}
 \caption[]{\small{A regression graph that is Markov equivalent to a covariance  graph. }}  \label{fig5}\vspace{-2mm}
 \end{figure}
 As discussed in the next section, to transfer such  equivalence results for graphs to a distribution generated over a regression graph, this distribution has to mimic
   some properties of joint Gaussian distributions; see also Wermuth and Sadeghi (2012).
   \subsubsection*{Implied independences in directed acyclic graphs\vspace{-2mm}}  
 
Work on reading all implied independences off a directed acyclic graph started with a path-braking, complex theorem, called d-separation; see  Pearl (1988), Geiger, Verma \& Pearl (1990). For the reduction of d-separation to the above  described  separation criterion for undirected graphs; see Lauritzen et  al.  (1990). 

A third criterion is based  on matrix representations of graphs, named their edge matrices. Square edge matrices  contain zeros for missing edges, ones for 
edges present and ones along the diagonal to extend the graph theoretic notion of  adjacency matrices in such a way that some sums or products of  edge matrices describe  transformations of graphs; see Wermuth, Wiedenbeck \& Cox (2006). It gives the most direct criterion in that it leads to a matrix  with the dimensions of two disjoint node subsets $\alpha$ and $\beta$: if this is a matrix of zeros only, then  it is established that the graph implies a desired conditional independence.

 In particular, let four disjoint subsets  $\alpha, \beta, c, m$  partition the node set of a directed acyclic graph,  let further $a=\alpha \cup m$, and 
$b=\beta \cup c$, then the transformation  to the edge matrix $\pcal_{a|b}$ parallels the transformation of parameters in  linear sequences of single-response regressions to the  matrix $\Pi_{a|b}$ 
of  least-squares regression coefficients with  $Y_a$ as  response and $Y_b$ as  regressor.
For joint Gaussian distributions, the submatrix of  $\Pi_{a|b}$ for rows $\alpha$ and columns $\beta$ is zero, $\Pi_{\alpha|\beta.c}=0$, if and only if $\alpha \ci \beta|c$. In general, $\pcal_{\alpha|\beta.c}$,  the corresponding submatrix of  $\pcal_{a|b}$ is zero  if  the given graph implies $\alpha \ci \beta |c$. 

For directed acyclic graphs, the three types of criteria have been proven to be equivalent; see Marchetti \& Wermuth (2009).   For combining independences in sequences of regressions, the generated distribution has to satisfy  additional conditions that mimic some properties of Gaussian distributions.

Independence properties of joint Gaussian  distributions have been nicely summarized in the information theory literature, together with other  properties that hold for all probability distributions;
see Ln\v{e}ni\v{c}ka and Mat\'u\v{s} (2007), Definition 1. To combine pairwise conditional independences in general sequences  regression just as in the Gaussian case, two properties are needed;  see Sadeghi  and Lauritzen (2014). We call the presence of the intersection property the `downward combination' and 
the presence of the composition property the `upward combination' of pairwise independences.

A graphical illustration of these two properties is given for just three nodes $i,h,k$ in Figure \ref{fig5a}.  In this simplest case,
upward and downward combination
mean, respectively, 
$$ (i\ci  h \text{ and } i\ci k) \implies  (i\ci h, k) \implies (i\ci h|k \text{ and } i\ci k|h),$$
$$ (i\ci  h|k  \text{ and } i\ci k|h) \implies  (i\ci h, k) \implies (i\ci h \text{ and } i\ci k) \, .$$
Essential for both properties in the above
    two equations is the first implication. The second implication holds
in all probability distributions and  just serves to motivate   the chosen names. For instance, upward combination means above to move from two simple to the two more complex  conditional independence statements; expressed more generally: it means moving  to increased conditioning sets.

On the left of Figure 6 is a complete regression graph with  joint response \{$h,k$\}, on the right  a complete directed acyclic graph with the single responses 
ordered as $(i, h, k)$
\begin{figure}[H]
\begin{center}\vspace{-7mm}
\includegraphics[scale=.46]{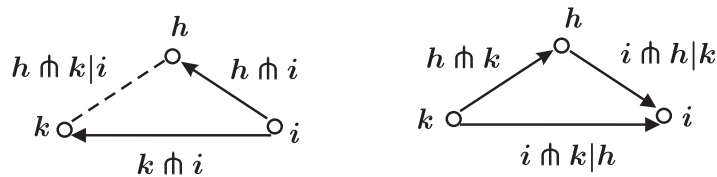}
\end{center}
\caption{\small Two regression graphs where removal of the $hi$-edge and the $ki$-edge leads to $i\ci h,k$, provided  
independences combine upward (left) and downward (right).}
\label{fig5a}
\end{figure}

A sufficient condition for the intersection property is to have a strictly positive  distribution; see San Martin, Mouchard \& Rolin (2005)  for extensions. Sufficient for the composition property is that
to every nonlinear effect there is also a linear one and that to every interactive effect there are also main effects. Such linear or main effects have to be substantial to be relevant
for a model fit; see N\'emeth and Rudas (2013) for an example
when this property does not hold in a sociological context.  In such  instances,  models more complex than graphical models are needed, such as  those included  for discrete variables in the  class of marginal models; see Bergsma and Rudas (2002).

\subsubsection*{Deciding on  implied dependences\vspace{-2mm}}

 Edge matrix results for directed acyclic graphs
have been extended to the more general types of regression graph and, more importantly, to inferring that a generating regression graph implies 
the conditional dependence $\alpha \pitchfork \beta|c$
if  $\pcal_{\alpha|\beta.c}\neq 0$; see Wermuth (2012).

To permit conclusions about dependences implied by sequences of regressions, the generated distribution has to have, in addition to upward and downward combination of independences,  non-vanishing parameters to each edge present in the  graph and it has to be dependence-inducing, a property also called singleton transitivity.  
Distributions with all   above mentioned properties have been called traceable,
since their regression graph  can be used to study pathways of development.

 In any specific application,  there may be special parametric constellations that lead to path cancellations and hence to preventing us from recognizing  dependences
that are implied by the given  generating process for other parametric constellations.  This cannot happen,
when all variables are coded, as in the Mannheim study, to have higher values for stronger adversities, for higher 
risks and for increased deficits and only linear dependences are 
observed or dependences that are accelerated compared to the linear effects.  In addition, prediction of the strength or  direction of induced associations  may vary with the measures of dependence used;
see Jiang, Ding \& Geng (2014 in press).

 For linear dependences among three variables,
there are already three types of possible parameters: covariances, concentrations
and regression coefficients. A detailed notation is needed  to understand their  relations:
\begin{equation}\sigma_{12|3}=\sigma_{12}- \sigma_{13}\sigma_{23}/\sigma_{33}, \label{ccov}\end{equation}
\begin{equation} \sigma^{23.1}= \sigma^{23}-   \sigma^{12}\sigma^{13},\label{mcon}\end{equation}
\begin{equation}\beta_{1|3}=\beta_{1|3.2}+\beta_{1|2.3}\beta_{2|3}, \label{pregc} \end{equation}
These three nice recursive properties were derived by different authors. Here,
 $\sigma_{12|3}$ denotes the conditional covariance of $1,2$ given $3$; see Anderson (1958), $\sigma^{23.1}$ 
 the concentration of $2,3$ marginalized over $1$; see Dempster (1969),  and  $\beta_{1|3.2}$ the population  
 coefficient of 2  when regressing 1 on both 2 and 3; see Cochran (1938) for the property.
 
 The right-hand sides of the above equations  contain, for trivariate   Gaussian distributions,  the parameters representing dependence  in different graphs for three nodes.  These parameters are in equation \eqref{ccov} for a covariance graph, in 
 \eqref{mcon} for a concentration graph and in  \eqref{pregc} for  a directed acyclic graph: with 1 as response to 2, 3 and 2 as a response to 3; see also Figure \ref{fig7}, last section.
 
 The  coefficient of 3 when regressing 1 only on 3, denoted by $\beta_{1|3}$  can be  regarded as the result
 of two paths: of $\beta_{1|3.2}$, an  arrow pointing from $3$ to $1$,  and of a sequence of two arrows  starting at $3$ and pointing to $1$ via 2 with  effect $\beta_{1|2
 .3}\beta_{2|3}$. Sometimes, the first is called the direct effect  of $3$ on $1$, the second the indirect effect and $\beta_{1|3}$ the overall effect.

If for instance $0=\sigma_{12}=\sigma_{13}$ in equation  \eqref{ccov}, then  $\sigma_{12|3}=0$  and consequently
also  $\beta_{1|3.2}=0$ in equation  \eqref{pregc} so that 1 has no dependence on 2 and  3 jointly. This is the Gaussian case of the more general notion of  combining independences upwards since here
$ (1\ci 2 \text{ and } 1 \ci 3) \implies 1\ci 2,3\,.$

With more variables, the  equations \eqref{ccov}  to \eqref{pregc} generalize, by adding everywhere a larger conditioning set in equations \eqref{ccov}  and \eqref{pregc}, to get, for instance,
$\sigma_{12|c}, \sigma_{12|3c}$ and $\beta_{1|3.c}, \beta_{1|3.2c}$, and by adding a larger marginalizing set in equation \ref{mcon}  to
get $\sigma^{23.m}, \sigma^{23.1m}$. The three measures of dependence relate  for a node set consisting of $1,2,3,c,m$ as
$$ \beta_{1|3.c}= \sigma_{13|c}/\sigma_{33|c}=-\sigma^{13.m}/\sigma^{11.m},$$
derived via the sweep operator  by Dempster (1969), and explaining  above why  $0=\sigma_{12|3}=\beta_{1|2.3}$ and $0=\sigma^{23.1}=\beta_{2|3}$ when all variances such as  $\sigma_{33|c}$ and all precisions, such as $ \sigma^{11.m}$, 
are nonzero in the two  single-response regressions.

For joint Gaussian distributions, the dependence-inducing property can be proven in its simplest form  using equation \eqref{ccov}: for $1 \pitchfork 3$ and $2 \pitchfork 3$ , one may have at most  $\sigma_{12|3}=0$ or $\sigma_{1|2}=0$ but never both. 
In the case of $\sigma_{12|3}=0$, the
induced correlation, $\rho_{12}=\sigma_{12}/\sqrt{\sigma_{11}\sigma_{22}}$, has also been named a `spurious correlation', since the 
dependence between the two variables 1 and 2 can be `explained away' by conditioning on their common neighbor 3.
 For three variables, the dependence-inducing property can be written as:
$$(1 \pitchfork 3 \text{ and }2\pitchfork  3) \implies \text{ at most } 1 \ci 2|3
\text{ or } 1\ci 2  \text{ but never both.}$$ 

It has been proven that binary variables are  dependence inducing; see Simpson (1951). But there are constellations of counts
for which both independences  $1 \ci 2|3$ and $1\ci 2$ seem to hold and  a decision would have to be based on outside information.

\subsubsection*{Independence-predicting and independence-preserving graphs\vspace{-2mm}}

The   results for deciding on implied  independences and dependences, based on an induced edge matrix, have been extended  to derive regression graphs 
implied by a given generating process  when, for instance, the order of the  responses or the conditioning sets or the marginalizing sets are changed;
see Wermuth \& Cox (2004).  

Then the  so-called independence-predicting 
graphs, may be derived by using the  partial closure operator  (Wermuth, Wiedenbeck \& Cox, 2006) as well as  sums and products of edge matrices, but each of these operations has also an intuitive  translation into
closing special types of paths in graphs; see Wermuth (2012).
 
For instance, when in a study different from the Mannheim one, the same variables were available except for cognitive deficits after 4 years, $Y_8, Y_4$, 
one predicts, from the
starting graph in Figure  \ref{fig3} and an
    unchanged order of the remaining variables, the regression graph in
    Figure  \ref{fig6}. In this case, it is just a subgraph of Figure  \ref{fig3} obtained 
     by removing nodes and edges of $Y_8$ and $Y_4$, no additional dependences are induced.

By contrast,  when $X_8$ and $ X_4$ are not available, all remaining variables turn out to be directly explanatory  for $Y_8$.
One consequence of this finding is that it is more important for  psychologists and psychiatrists to evaluate also motor development than it is for physicians to take also  cognitive development into account.

\begin{figure}[H]\vspace{-4mm}
\begin{center}
\includegraphics[scale=.51]{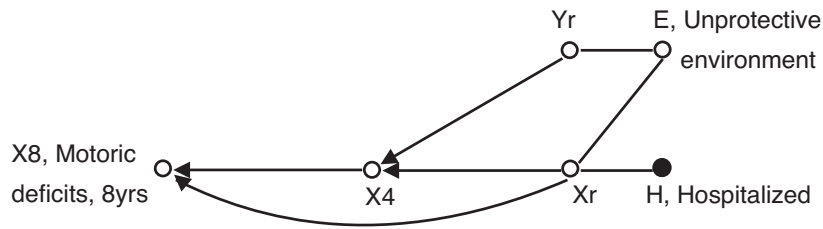}
\end{center}\vspace{-4mm}
 \caption[]{\small{The regression graph induced by Figure \ref{fig3} after marginalizing over $Y_8$ and $Y_4$.}}
 \label{fig6}
\end{figure}

Typically, one cannot use such an induced regression graph to condition on or marginalize over more nodes and still see what the
starting graph would have implied for this case, while this is possible with the so-called independence-preserving graphs.  
Starting from directed acyclic graphs, three types of such graphs have been proposed, the MC-graphs by Koster (2002), the maximal ancestral graphs by 
 Richardson and Spirtes (2002)  and the summary graphs by Wermuth (2011). For their relations  and proofs of Markov equivalence,  
 see Sadeghi (2013). To construct these types of graph within the program environment R, see Sadeghi and Marchetti  (2012).
 
 Summary graphs may be derived for the  more general regression graphs and, if obtained after marginalizing  only, they provide graphical criteria to detect  confounding.
  
 \subsubsection*{Graphical representations of distortions\vspace{-2mm}}
For traceable regressions, the regression graph represents  an independence structure and a partly specified dependence structure. The graph can be viewed as  a hypothesis  of how sequences of single or joint responses 
generate a joint distribution when, for instance, corresponding point estimates of parameters are taken as the 
population values. For  reliable interpretations of estimated effects, it is  important to prevent, if at all possible, any important, sizeable distortion of   the actual population parameters.

Different types of distortions of a treatment effect may  be avoided by using  regression graphs and appropriate 
study designs. To see this, we show in Figure \ref{fig7} simple cases  of   under-conditioning, of over-conditioning and 
of direct confounding by using  a node crossed by two lines, $\margn$,  to indicate that the variable is to be marginalized over and a node surrounded by a square, $\condnc$, to indicate that the variable is to be conditioned on.

For instance, by using label 3 for treatment and label 1 for the outcome or response to 2 and 3, the graph on the left of Figure \ref{fig7}
shows one important intermediate variable, labelled 2, which could be, say,  compliance of the patients
to treatment.  

\begin{figure}[H]
\begin{center}\vspace{-4mm}
{\n \includegraphics[scale=.56]{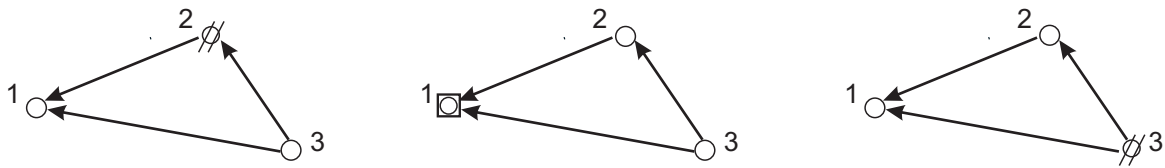}}
\end{center}\vspace{-4mm}
 \caption[]{\small{Left: under-conditioning for $1\pitchfork 3|2$ by ignoring an  intermediate variable, middle: over-conditioning for $2\pitchfork3$ by selecting levels  of the common response 1 and right:
 direct confounding for $1\pitchfork 2|3$ by ignoring the common explanatory variable 3 for 1 and 2; generating 
 a double edge  $1 \fladal\; 2$  in the summary graph obtained by marginalizing over node 3.}}
\label{fig7}
\end{figure}

In particular, when differential compliance is ignored, a distorted
overall effect results
which then coincides with the result of a so-called intention-to-treat analysis. Conditioning on the common response in the middle graph distorts the simple dependence  of 2 on 3 and is also named a selection bias. 
More complex cases of such over-conditioning may occur, for instance, when there is an effect for $ 2 \fla 3$ and, in addition,  a path like $2 \fra \condnc \fla \margn \fra  \condnc \fla 3.$

The simplest case of direct confounding, shown on the right of Figure \ref{fig7}, is also called the presence of an
unmeasured confounder. It has a longer history than graphical models; see Vandenbroucke (2002). It is avoided
when there is a successful, fully randomized allocation of individuals to treatment levels, since in that case, all effects on treatment, observed or unobserved, are removed. This design leads also to a removal of all incoming arrows to the treatment  variable in a regression graph and to the absence of any double edge
 in corresponding summary graphs,  where the only possible double edge  is  $\snode \fladal\;\snode$.

Another source of distortion, named indirect confounding, may lead to strong dependence reversal compared to the population dependence; see Wermuth and Cox (2008). A first  example is due to
Robins and Wasserman (1997), shown here in Figure \ref{fig8}, left 
and the summary 
    graph after marginalizing over U in Figure \ref{fig8}, right

Notice that in this  regression graph there are no unmeasured confounders of $Y \fla T_p$ and there
is no over-conditioning and no under-conditioning for this dependence of $Y$ on $T_p$. But $A$ is  intermediate between $Y$ and $T_p$. By conditioning on just the  regressors $T_r$ and $T_p$ of $Y$ that remain when $U$ is unobserved, one conditions implicitly also
on their past, hence on $A$ and thereby activates the path $Y\dal A\fla X_p$ to distort  the conditional dependence $Y \pitchfork T_p| T_r, U$ as it is present in the generating graph.  

\begin{figure}[H]\vspace{-4mm}
\begin{center}
{\n \hspace{-3.5cm} \includegraphics[scale=.56]{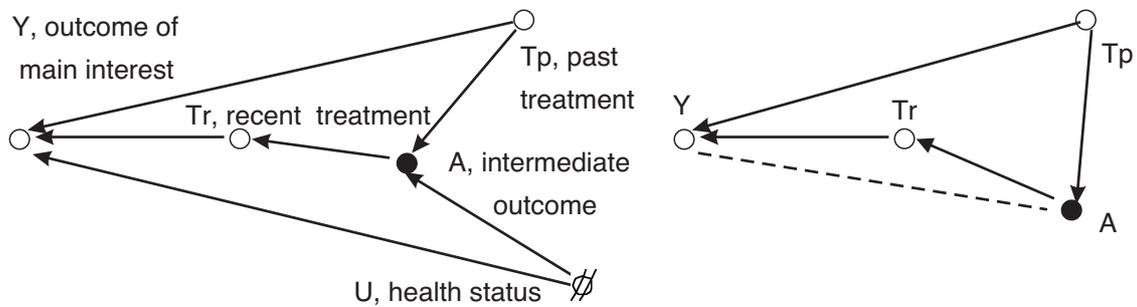}}
\vspace{-2mm}
\end{center}
 \caption[]{\small{Left: full randomization for $T_p$ leads to removal of $T_p \fla U$, randomized allocation of individuals to selected levels of  a more recent treatment, $T_r$, given  results of the intermediate outcome $A$, leads to removal of $T_r\fla U$ and of $T_r \fla T_p$ and  to the presence of $T_r \fla A$; right: the summary graph obtained by marginalizing over   $U$;  path $Y\dal A\fla T_p$ explaining the indirect confounding of $Y\fla T_p$ when response $Y$ is regressed on only  $T_p$ and $T_r$.} }
\label{fig8}
\end{figure}

More
generally, indirect confounding will result for  $Y\fla T$ in a regression graph, when  in the summary graph obtained after marginalizing over all unobserved
variables, one or more of  the following two types of paths  are present; see Wermuth (2011):
$$ Y\dal \snode \dal \snode \ldots \snode \dal \snode \dal T\nn  \text{ or } \nn Y\dal \snode \dal \snode \ldots \snode \dal \snode \fla T ,$$
where every inner node along the path is an intermediate variable between $Y$ and $T$ and   $\ldots $ denotes a possible continuation of the same type of neighboring edges.

 Indirect confounding, which can be much stronger than direct confounding, appears to have been  largely
 ignored  so far  in the literature, not only in the statistical one, but also in the  current literature on causal modeling, based on virtual interventions using directed acyclic graphs or structural equations; see Pearl (2014 in press), Richardson and Robins (2013). Only direct confounding and
 selection bias are frequently considered.
 
 The specification of  generating processes via sequences of single or joint response regressions    lead
 to regression graphs that may be needed for useful causal interpretations and to corresponding summary graphs that help to avoid possible mistaken interpretations  
 when some variables are hidden, that is latent or unobserved, and some responses are conditioned on.

\subsubsection*{Bibliography\\[-4mm]}
\begingroup
\renewcommand{\section}[2]{}
    
 \endgroup

\end{document}